\documentstyle[11pt,newpasp,twoside,epsf]{article}
\def\@journalname{ASP Conference Series}
\def\cpr@holder{Astronomical Society of the Pacific}
\def\@jourvol{**VOLUME***}
\def\cpr@year{2003}
\def\vol@title{Magnetic Cataclysmic Variables}
\def\vol@author{M. Cropper \& S. Vrielmann}

\def\edcomment#1{\iffalse\marginpar{\raggedright\sl#1\/}\else\relax\fi}
\reversemarginpar

\begin{document}
\title{The {\it Chandra} Observation of the IP TX Col}
 \author{Eric M. Schlegel}
\affil{Harvard-Smithsonian Center for Astrophysics,  60 Garden St.,
 Cambridge, MA 02138 USA}
\author{Anandi Salinas}
\affil{University of Texas at Austin, Austin, TX 78712 USA}

\begin{abstract}
We present a preliminary look at the serendipitous observation of the
intermediate polar TX Col by {\it Chandra}.  The $\sim$52 ksec
observation is uninterrupted, providing an opportunity to disentangle
the light curve and power spectra components.  We illustrate the
energy-dependence of the power spectrum.
\end{abstract}

\section{Introduction}

TX Col is a 15-th mag object discovered by HEAO-1 (Wood et al. 1984)
and recognized as an intermediate polar in 1985/6 (Buckley et
al. 1985; Tuohy et al. 1986).  Its orbital period is 20,592 sec with a
spin period of $\sim$1910 sec (Tuohy et al 1986; Mhlahlo et al., these
proceedings).

\section{{\it Chandra} Observation}

TX Col was observed by {\it Chandra} on 2000 July 26/27 for $\sim$53 ksec
(observation identification number 914); the target was the cluster of
galaxies Cl J0542.8-4100 to measure the X-ray temperature function of
clusters (e.g., Ikebe et al. 2002).  The cluster was within 1$'$ of
the aimpoint, placing TX Col $\sim$6$'$.7 arcmin off-axis on CCD 1 of
the `imaging' portion of the ACIS detector.  Figure 1 shows the {\it
Chandra} field-of-view showing that 5 ACIS CCDs were active during the
observation.  TX Col and the galaxy cluster are indicated.

The estimated count rate was $\sim$0.57 counts s$^{-1}$ with a
background rate of 0.006 counts s$^{-1}$.  Both values were obtained
by extracting the counts in circles of radii 0.3 and 1 arcmin,
respectively.  The background circle was displaced from the position
of TX Col by $\sim$2$'$ because TX Col falls sufficiently close to the
edge of the CCD that $\sim$half of an annulus centered on TX Col's
position would have extended beyond the chip's edge.  A narrow
rectangular region was excluded from the outer portions of the TX Col
extraction to eliminate the presence of the frame transfer events.

\begin{figure}
\caption{{\it Chandra} ACIS field-of-view of TX Col.}
\plotfiddle{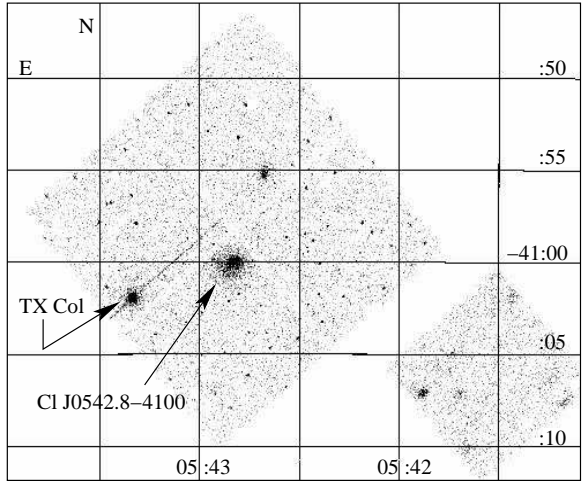}{3.0in}{-90}{50}{50}{-200}{250}
\end{figure}

The off-axis angle reduces the effects of event pileup (Davis 2001).
The streak caused by the frame transfer determines the position of TX
Col along the CCD columns.  We adopted the peak number of events on
that line as the center of the observed point spread function.  We
extracted a radial profile using 21 elliptical annuli spread across
$\sim$21 pixels ($\sim$10$''$.5); a theoretical point spread function
was generated from the {\it Chandra} PSF library at the observed
off-axis angle.  The two functions agreed to better than 5\% in the
core of the profile and indicated that pileup is at most a minor
component of the TX Col observation.

\begin{figure}
\caption{Energy-filtered light curves folded at the spin period of TX
Col.}
\plotone{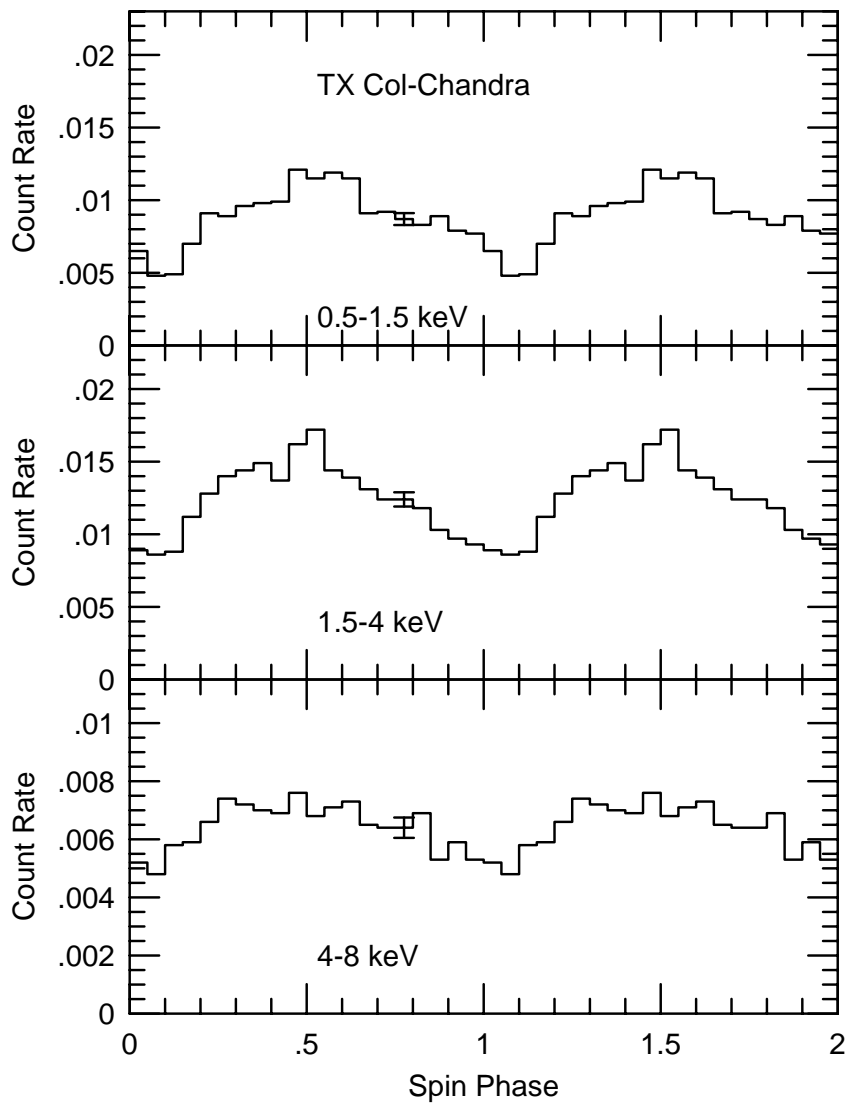}
\end{figure}

\section{Light Curves}

We filtered the events into three energy bands: 0.5-1.5, 1.5-4, and
4-8 keV in an effort to separate the high energy source emission and
the low-energy absorption.  The middle band served simply to permit
defining a color.  We extracted the events and binned them into 20
bins folded on the spin period of 1910 sec.  We arbitrarily adopted
the time of the first {\it Chandra} event as the starting point for
the phasing for this preliminary analysis.  Figure 2 shows the result.
The zero point is incorrect by $\sim$0.5 in phase (depending upon how
one defines the `peak').  The soft band shows a humped behavior
approximately symmetric about the maximum.  The medium band shows a
steeper rise to maximum with a gradual fall.  The hard band is
flattened with perhaps a steep rise just after phase 0.0.

\begin{figure}
\caption{The complete {\it Chandra} ACIS light curve.  About 26 individual
spin cycles are visible in this observation.  Note the amplitude
variations across the observation. (top) the full observation;
(bottom) a portion expanded.  The bins are 100 sec in size.}
\plotone{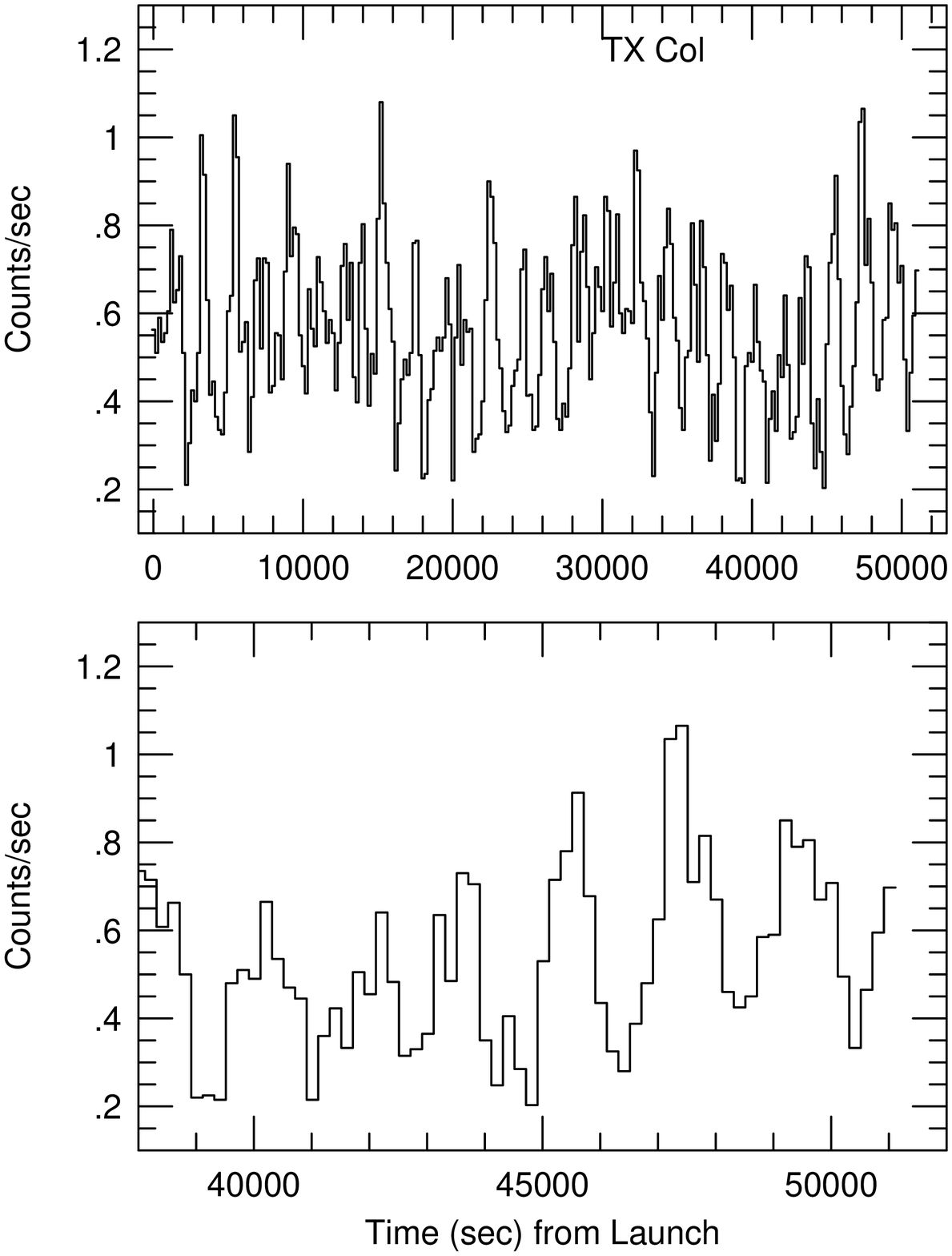}
\end{figure}

Figure 3 (top) shows the complete ACIS light curve, $\sim$52 ksec or
$\sim$26 spin cycles in length.  We can easily count the spin cycles,
but note that the envelope of the amplitudes changes significantly
across the observation.  Figure 3 (bottom) expands a portion of the
full light curve to show the significant changes in amplitude among
individual spin cycles.

\section{Power Spectra}

\begin{figure}
\caption{Energy-filtered power spectra of TX Col using the three
energy band definitions indicated on the figure.  The vertical dashed
lines indicate the positions of the $|m{\omega}{\pm}n{\Omega}|$ for
(m, n) = (0, 1, 2).  Note that the vertical scale changes for the 4-8
keV band.  Also note the dominance of the ${\omega}-{\Omega}$
component in that band.}
\plotone{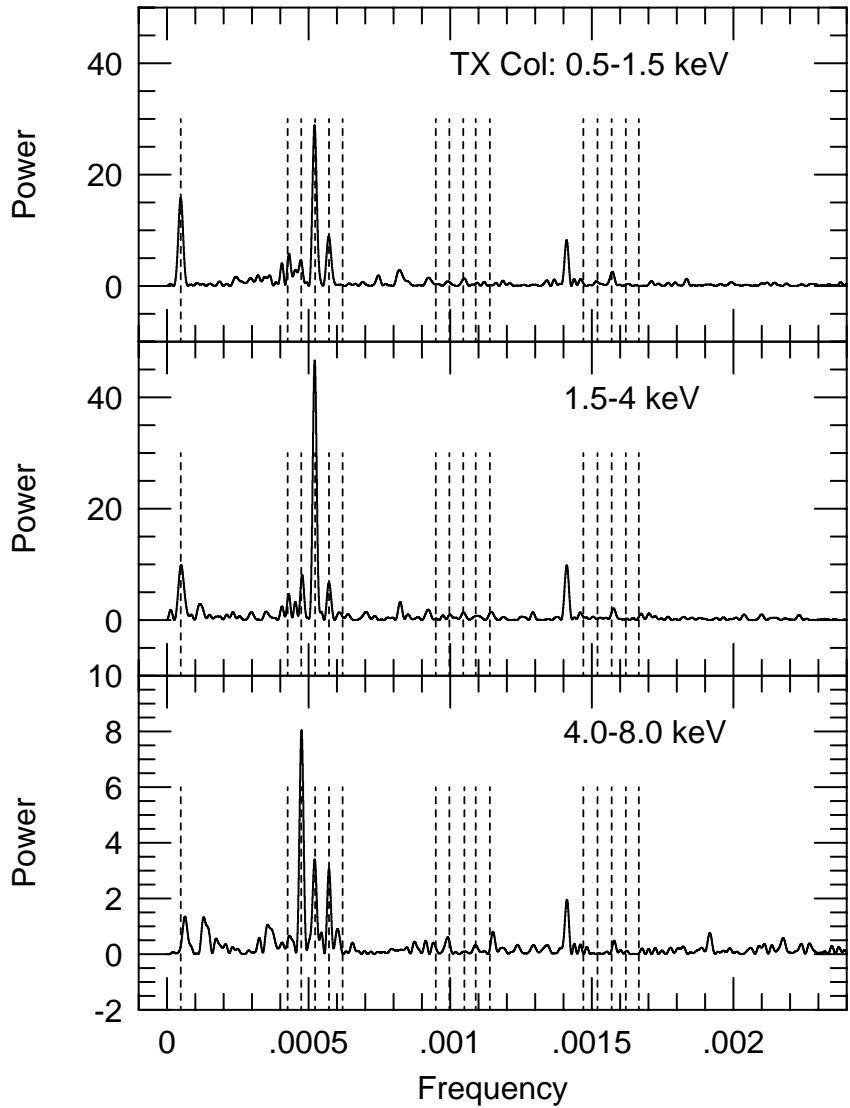}
\end{figure}
 
Power spectra were computed in the three energy bands for the entire
observation as well as across the entire ACIS bandpass for subsets of
the observation.  There are insufficient events to examine power
spectra of energy-filtered subsets.

\begin{figure}
\caption{Power spectrum from the (top) first `orbit' and from the
(bottom) `second' orbit.  Note that the ${\omega}-{\Omega}$ component
is essentially missing from the second orbit.  This figure replaces
one shown at the meeting (the `half-orbit' approach), but the results
are insignificantly different.}  
\plotone{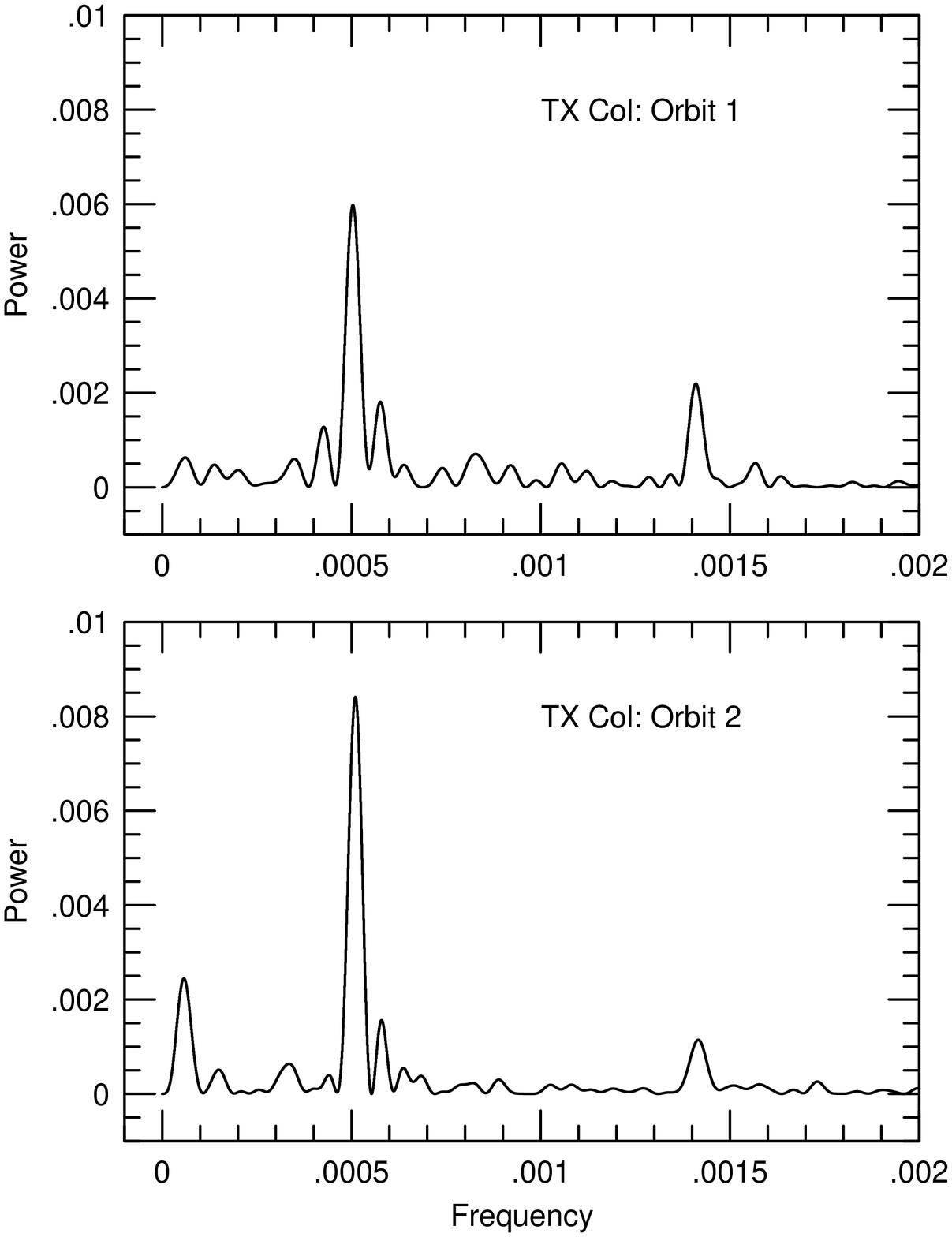}
\end{figure}

The power spectra illustrate differences in amplitudes of the various
components as a function of energy.  The spin frequency is the
dominant component in the soft and medium bands, but at the highest
energies, the ${\omega}-{\Omega}$ component dominates.  Also visible
is significant power at the half-integer frequencies, particularly
near 8.2$\times$10$^{-4}$ Hz, suggesting the presence of a
2${\omega}-{9 \over 2}{\Omega}$ component.  

If we split the observation into two `orbits' at precisely the orbital
period (the first 20592 sec and the second), and Fourier transform the
data, we obtain the power spectra shown in Figure 5.  Note that the
${\omega}-{\Omega}$ component is essentially missing from the second
orbit.

Table 1 shows a history of the detected periods from {\it ASCA}
and {\it ROSAT} (Norton et al. 1997), and the {\it Chandra}
observation.  Additional periods may be uncovered in the {\it Chandra}
observation as the analyses of the data are refined.  Table 1
represents the primary results to date; quantitative measurements will
be included in the final paper describing the {\it Chandra}
observation; that paper is expected to be submitted for publication in
early spring 2003.  At this point, the power spectra imply the soft
X-rays are disk-fed because the amplitude of the ${\omega}-{\Omega}$
component is approximately equal to the amplitude of the
${\omega}+{\Omega}$.  Our goal with this observation is the Fourier
decomposition of the light curve (Warner 1986; Norton et al. 1996).

\acknowledgments

Support for AS was provided by the US National Science Foundation
through the Research Experience for Undergraduates program.  The
research of EMS was supported by NASA Contract NAS8-39073 to SAO for
the {\it Chandra} Observatory.

\begin{table}
\caption{Significant X-ray Periods: History}
\begin{tabular}{|c|c|c|c|c|c|c|c|c|} 
\tableline
Satellite & Date & ${\Omega}$ & ${\omega}$ & ${\omega}-{\Omega}$ &
${\omega}+{\Omega}$ & ${\omega}-2{\Omega}$ & 2${\omega}-2{\Omega}$ &
2${\omega}$ \\
\tableline
{\it ASCA} & 10/1994 & ${\surd}$ & ${\surd}$ & ${\times}$ & ${\times}$ &
${\times}$ & $\surd$ & $?$ \\
{\it ROSAT}& 10/1995 & ${\surd}$ & ${\surd}$ & ${\surd}$ & ${\times}$ &
${\surd}$ & $\times$ & ${\surd}$ \\
{\it Chandra} & 7/2000 & ${\surd}$ & ${\surd}$ & ${\surd}$ & $\surd$ &
$\surd$ & ${\surd}$ & ${\surd}$ \\ 
\tableline \tableline
\end{tabular}
\end{table}

\end{document}